\documentclass[final,5p,times,twocolumn,nopreprintline]{elsarticle}
\usepackage{amsmath,slashed,booktabs}
\usepackage{graphicx,graphics}
\usepackage{dcolumn}
\usepackage[hyperfootnotes=false]{hyperref}
\usepackage{xspace}
\usepackage{color}
\usepackage{balance}
\usepackage{multirow}

\usepackage{fancyhdr}
\fancypagestyle{firstpage}{%
	
}
\newcommand{\GeV}{\,\text{GeV}}
\newcommand{\MeV}{\,\text{MeV}}

\newcommand{\fm}{\,\text{fm}}

\newcommand{\mpi}{M_\pi}

\renewcommand{\Im}{\text{Im}\,}

\newcommand{\beq}{\begin{equation}}
\newcommand{\eeq}{\end{equation}}

\allowdisplaybreaks[1]

\begin{document}

\renewcommand{\theequation}{\arabic{equation}}

\begin{frontmatter}

\title{Chiral extrapolation of hadronic vacuum polarization}   

\author[Bern]{Gilberto Colangelo}
\author[Bern]{Martin Hoferichter}
\author[Bonn]{Bastian Kubis}
\author[Bonn]{Malwin Niehus}
\author[Bern]{Jacobo Ruiz de Elvira}

\address[Bern]{Albert Einstein Center for Fundamental Physics, Institute for Theoretical Physics, University of Bern, Sidlerstrasse 5, 3012 Bern, Switzerland}
\address[Bonn]{Helmholtz-Institut f\"ur Strahlen- und Kernphysik (Theorie) and
Bethe Center for Theoretical Physics, Universit\"at Bonn, 53115 Bonn, Germany}

\begin{abstract}
  We study the pion-mass dependence of the two-pion channel in the
  hadronic-vacuum-polarization (HVP) contribution to the anomalous magnetic
  moment of the muon $a_\mu^\text{HVP}$, by using an Omn\`es representation
  for the pion vector form factor with the phase shift derived from the
  inverse-amplitude method (IAM). Our results constrain the dominant
  isospin-$1$ part of the isospin-symmetric light-quark contribution, and
  should thus allow one to better control the chiral extrapolation of
  $a_\mu^\text{HVP}$, required for lattice-QCD calculations performed at
  larger-than-physical pion masses. In particular, the comparison of the
  one- and two-loop IAM allows us to estimate the associated systematic 
  uncertainties and show that these are under good control.
\end{abstract}

\end{frontmatter}

\thispagestyle{firstpage}

\section{Introduction}

Currently the biggest uncertainty in the Standard-Model prediction for the anomalous magnetic moment of the muon~\cite{Aoyama:2020ynm,Aoyama:2012wk,Aoyama:2019ryr,Czarnecki:2002nt,Gnendiger:2013pva,Davier:2017zfy,Keshavarzi:2018mgv,Colangelo:2018mtw,Hoferichter:2019gzf,Davier:2019can,Keshavarzi:2019abf,Hoid:2020xjs,Kurz:2014wya,Melnikov:2003xd,Colangelo:2014dfa,Colangelo:2014pva,Colangelo:2015ama,Masjuan:2017tvw,Colangelo:2017qdm,Colangelo:2017fiz,Hoferichter:2018dmo,Hoferichter:2018kwz,Gerardin:2019vio,Bijnens:2019ghy,Colangelo:2019lpu,Colangelo:2019uex,Blum:2019ugy,Colangelo:2014qya}
\beq
\label{amuSM}
a_\mu^\text{SM}=116\,591\,810(43)\times 10^{-11} 
\eeq
resides in the HVP contribution, which, when derived from $e^+e^-\to\text{hadrons}$ cross-section data~\cite{Aoyama:2020ynm,Davier:2017zfy,Keshavarzi:2018mgv,Colangelo:2018mtw,Hoferichter:2019gzf,Davier:2019can,Keshavarzi:2019abf,Hoid:2020xjs}
\beq
\label{ee}
a_\mu^\text{HVP}\big|_{e^+e^-}=6\,931(40)\times 10^{-11},  
\eeq
leads to a $4.2\sigma$ difference to experiment~\cite{bennett:2006fi,Abi:2021gix,Albahri:2021ixb,Albahri:2021kmg,Albahri:2021mtf}
\beq
\label{exp}
a_\mu^\text{exp}=116\,592\,061(41)\times 10^{-11}.
\eeq
Improving the (time-like) data-driven evaluation of HVP~\eqref{ee} relies on new data, most crucially for the $e^+e^-\to 2\pi$ channel~\cite{SND:2020nwa,Ignatov:2019omb}, while a space-like measurement would be possible at the 
MUonE experiment~\cite{Abbiendi:2016xup,Banerjee:2020tdt}.

Alternatively, the precision of the HVP contribution evaluated in lattice QCD is getting closer to the data-driven determination, with an average~\cite{Aoyama:2020ynm} (based on
Refs.~\cite{Chakraborty:2017tqp,Borsanyi:2017zdw,Blum:2018mom,Giusti:2019xct,Shintani:2019wai,Davies:2019efs,Gerardin:2019rua,Aubin:2019usy,Giusti:2019hkz})
\beq
\label{eq:LatticeAverage}
a_\mu^\text{HVP}\big|_\text{lattice\,average}=7\,116(184)\times 10^{-11},
\eeq
and a subsequent first sub-percent result~\cite{Borsanyi:2020mff}
\beq
a_\mu^\text{HVP}=7075(55)\times 10^{-11}.
\eeq 
In this Letter, we do not address the $2.1\sigma$ tension with the data-driven approach,\footnote{In contrast, there is good agreement between data-driven and lattice-QCD evaluations for hadronic light-by-light scattering, as further corroborated by recent work~\cite{Hoferichter:2020lap,Ludtke:2020moa,Bijnens:2020xnl,Bijnens:2021jqo,Zanke:2021wiq,Chao:2021tvp,Danilkin:2021icn,Colangelo:2021nkr}.} see Refs.~\cite{Lehner:2020crt,Crivellin:2020zul,Keshavarzi:2020bfy,Malaescu:2020zuc,Colangelo:2020lcg}, but instead focus on the potential source of systematic uncertainty in lattice calculations that may arise if the simulation is performed at unphysical values of the quark masses. 

This effect is most relevant for the isospin-symmetric $ud$ correlator,
both because its contribution is by far the largest, and because it is the
lightest quarks that make simulations at the physical point expensive. Often,
the required quark-mass extrapolation can be controlled using chiral
perturbation theory (ChPT), at least for sufficiently small masses, but the
analysis of Ref.~\cite{Golterman:2017njs} showed that for the HVP
contribution this does not seem to be the case. 
On the one hand, the presence of a mass scale lighter
than $\mpi$, namely the muon mass, makes the pure chiral expansion of practical use only for $\mpi \ll m_\mu$~\cite{Golterman:2017njs}.
Physically, it is well
known that the $2 \pi$ contribution to HVP is dominated by the $\rho(770)$
meson, see, e.g., Ref.~\cite{Erben:2019nmx} for the implication for lattice
calculations, and that controlling the quark-mass dependence of its
parameters requires information beyond ChPT. On the other hand, one would
not expect the quark-mass dependence of the $\rho(770)$ mass, for example,
to be so complicated that it could not be described by a simple
parameterization. If this is the case, it is not clear why a simple
parameterization of the quark-mass dependence of the $2 \pi$ contribution
to HVP should not be possible, and even allow for a controlled chiral
extrapolation of good precision (in fact, finite-volume corrections have been addressed using ChPT methods~\cite{Aubin:2020scy}). Given the high computational cost of
simulations at the physical quark masses this is a question of current
high interest, which can be addressed from a ChPT/phenomenological point of
view and deserves the detailed investigation we aim to provide in this
Letter.

Our approach here is to follow Ref.~\cite{Guo:2008nc} and combine an
Omn\`es description~\cite{Omnes:1958hv} of the pion vector form factor
(VFF) with the inverse-amplitude method
(IAM)~\cite{Truong:1988zp,Dobado:1989qm,Truong:1991gv,Dobado:1992ha,Dobado:1996ps,Guerrero:1998ei,GomezNicola:2001as,Nieves:2001de},
to capture the quark-mass dependence of the dominant two-pion intermediate
states. To this end, we employ the one- and two-loop IAM to describe the
pion-mass dependence of the $\pi\pi$ $P$-wave phase
shift~\cite{Niehus:2020gmf}, leading to a representation that guides the
chiral extrapolation of the $I=1$ component of the isospin-symmetric $ud$
contribution to $a_\mu^\text{HVP}$. We stress that our goal is not to show
that the IAM is able to {\em predict} to high precision the quark-mass dependence of
the $\pi\pi$ $P$-wave phase shift, but rather whether it is able to
{\em describe} it, and we trust that the analysis in Ref.~\cite{Niehus:2020gmf}
provides a positive answer to this question.

$I=0$ and isospin-breaking terms are much smaller in size, in such a way
that the systematic uncertainty in their extrapolation becomes less
critical. Further, effects from inelastic states (mainly $4\pi$) are
sufficiently small that standard polynomial extrapolations should be
sufficient.  For the dominant $2\pi$ contribution, which we can capture
with the IAM, the comparison of one- and two-loop extrapolations provides a
measure of the systematic uncertainty, and thus allows the complete
quantification of uncertainties that arise when ensembles at
heavier-than-physical pion masses are included in the analysis.

We establish the formalism in Sec.~\ref{sec:formalism}, reviewing the relevant aspects of HVP,  Omn\`es methods for the pion VFF, and the IAM. For the applications described in Sec.~\ref{sec:extrapolation}, we will first impose the physical point from data and show the resulting quark-mass dependence of the HVP integral, before turning to strategies how the corresponding constraints could be implemented in lattice analyses in Secs.~\ref{sec:spacelike} and \ref{sec:strategies}. We conclude in Sec.~\ref{sec:conclusions}.

\section{Formalism}
\label{sec:formalism}

In the data-driven approach the HVP contribution is calculated as~\cite{Bouchiat:1961lbg,Brodsky:1967sr}
\begin{align}
\label{amu_HVP}
 a_\mu^\text{HVP}&=\bigg(\frac{\alpha m_\mu}{3\pi}\bigg)^2\int_{s_\text{thr}}^\infty ds \frac{\hat K(s)}{s^2}R_\text{had}(s),\notag\\
 R_\text{had}(s)&=\frac{3s}{4\pi\alpha^2}\sigma(e^+e^-\to\text{hadrons}),
\end{align}
where $\hat K(s)$ is a known kernel function and the hadronic cross section is photon inclusive. In contrast, lattice QCD does not proceed via the $R$-ratio, but instead employs a representation~\cite{Lautrup:1971jf,Blum:2002ii,Bernecker:2011gh}
\begin{align}
 a_\mu^\text{HVP}&=\bigg(\frac{\alpha}{\pi}\bigg)^2\int_{0}^\infty dt\, \tilde K(t) G(t),
 \label{timemomentum}
\end{align}
where $\tilde K(t)$ is another analytically known kernel function and $G(t)$ is determined by the correlator of two electromagnetic currents $j_\mu^\text{em}$
\begin{align}
 G(t)&=-\frac{a^3}{3}\sum_{k=1}^3\sum_{\mathbf x}G_{kk}(t,{\mathbf x}),\notag\\
 G_{\mu\nu}(x)&=\langle 0|j_\mu^\text{em}(x)j_\nu^\text{em}(0)|0\rangle,
\end{align}
where $a$ is the lattice spacing and the limit $a\to 0$ implied in the end. 
This shows that in this approach the contributions of particular channels
in $R_\text{had}(s)$ cannot be resolved, while instead the calculation is
organized in a flavor decomposition, separated into an isospin-symmetric
$ud$ contribution, other quarks flavor-by-flavor, and isospin-breaking
corrections (both electromagnetic and from the quark-mass difference
$m_u-m_d$). For that reason, information on the quark-mass dependence of a
particular hadronic channel, in general, does not translate to an
extrapolation prescription for lattice calculations. However, the $I=1$
component of the isospin-symmetric $ud$ correlator does correspond
predominantly to two-pion intermediate states, with effects from other
possible states, such as $4\pi$, appreciably suppressed. Since, in
addition, the lightest states are expected to be most affected by
non-trivial features of the chiral extrapolation, we will assume that such
subleading effects can be adequately described by a polynomial, with the
chiral behavior of $a_\mu^\text{HVP}[\pi\pi]$ thus a proxy for that of
$a_\mu^\text{HVP}[ud, I=1]$.

The two-pion contribution to $R_\text{had}(s)$ can be expressed in terms of the pion VFF $F_\pi^V(s)$
\beq
\label{VFF_def}
	\langle \pi^\pm(p') | j_\text{em}^\mu(0) | \pi^\pm(p) \rangle =\pm (p'+p)^\mu F_\pi^V((p'-p)^2),
\eeq
with
\beq
	\sigma(e^+e^-\to\pi^+\pi^-) = \frac{\pi \alpha^2}{3s} \sigma_\pi^3(s) \big| F_\pi^V(s) \big|^2 ,
\eeq
and $\sigma_\pi(s) = \sqrt{1-4\mpi^2/s}$.
$F_\pi^V(s)$ is then strongly constrained by $\pi\pi$ scattering, as reflected by the fact that up to a polynomial the combination of analyticity and unitarity equates the elastic contributions to the VFF with the Omn\`es factor~\cite{Omnes:1958hv}
\beq
\label{omnes}
\Omega_1^1(s) = \exp\left\{ \frac{s}{\pi} \int_{4\mpi^2}^\infty ds^\prime \frac{\delta_1^1(s^\prime)}{s^\prime(s^\prime-s)} \right\},
\eeq
where $\delta_1^1(s)$ is the $P$-wave $\pi\pi$ scattering phase shift. This connection between the VFF and $\pi\pi$ scattering has been employed in numerous works in the literature, see, e.g., 
Refs.~\cite{DeTroconiz:2001rip,Leutwyler:2002hm,Colangelo:2003yw,deTroconiz:2004yzs,Hoferichter:2016duk,Hanhart:2016pcd,Ananthanarayan:2018nyx,Hoferichter:2018zwu,Ananthanarayan:2020vum}, and also forms the basis for the present analysis. In general, inelastic and isospin-breaking corrections need to be considered for a phenomenologically viable description~\cite{Leutwyler:2002hm,Colangelo:2003yw,Colangelo:2018mtw}   
\beq
	\label{eq:PionVFF}
	F_\pi^V(s) = \Omega_1^1(s) G_\omega(s) G_\text{in}(s),
\eeq
including factors $G_\omega(s)$ and $G_\text{in}(s)$ that account for
$3\pi$ and $4\pi$ intermediate states, respectively, with the former
dominated by $\rho$--$\omega$ mixing and the latter by the $\omega \pi$
channel, which justifies the expansion in a conformal polynomial.

For the quark-mass extrapolation of $a_\mu^\text{HVP}[ud]$ this
representation can be simplified in several ways. First, since
isospin-breaking effects are booked elsewhere in lattice calculations, we
can set $G_\omega(s)=1$ and ignore final-state-radiation corrections to the
cross section. The effects of inelastic states on the $2 \pi$-channel below
$1\GeV$ are small and well described by a conformal polynomial of low
degree~\cite{Colangelo:2018mtw}: we will truncate the integral in
Eq.~\eqref{amu_HVP} at $\Lambda=1\GeV$ (with the threshold at
$s_\text{thr}=4\mpi^2$). In order to simplify the analysis of its quark-mass dependence we will first replace $G_\text{in}(s)$ by a polynomial and
consider its coefficients as parameters in the lattice analysis, meant to
subsume inelastic effects. For a linear polynomial, $G_\text{in}(s)=1+\beta
s$, the free parameter is related to the pion charge radius via \beq
\label{beta}
\langle r_\pi^2\rangle=6\frac{d F_\pi^V(s)}{d s}\bigg|_{s=0} =6\big[\beta +
\dot\Omega^1_1(0)\big], \eeq where $\dot\Omega^1_1(0)$ denotes the
derivative of the Omn\`es factor at $s=0$. At the physical point all
parameters are then determined via Eqs.~\eqref{omnes} and~\eqref{beta},
using input for $\delta_1^1(s)$ and $\langle r_\pi^2\rangle$ from
Ref.~\cite{Colangelo:2018mtw} (derived from a fit to the data sets of
Refs.~\cite{Amendolia:1986wj,Akhmetshin:2001ig,Akhmetshin:2003zn,Achasov:2005rg,Achasov:2006vp,Akhmetshin:2006wh,Akhmetshin:2006bx,Ambrosino:2008aa,Aubert:2009ad,Ambrosino:2010bv,Lees:2012cj,Babusci:2012rp,BESIII:2015equ,Anastasi:2017eio},
including constraints from $\pi\pi$ Roy
equations~\cite{Roy:1971tc,Ananthanarayan:2000ht,GarciaMartin:2011cn,Caprini:2011ky}
and the Eidelman--\L{}ukaszuk
bound~\cite{Lukaszuk:1973jd,Eidelman:2003uh}). The final representation for
the VFF in the isospin limit then reads \beq
\label{VFF_isospin_limit}
F_\pi^V(s)\big|_{\epsilon_\omega=0} = \bigg[1+\bigg(\frac{\langle
  r_\pi^2\rangle}{6}-\dot\Omega^1_1(0)\bigg)s\bigg]\Omega_1^1(s), \eeq
where, by including information on the charge radius, we have incorporated
the dominant inelastic effects. We will show below that the switch from a
polynomial in $s$ to one in a conformal variable does not change the
results of our analysis.

The quark-mass dependence of the resulting $a_\mu^\text{HVP}[\pi\pi, \leq1\GeV]$ is taken from the IAM, using the analytic expressions from Ref.~\cite{Niehus:2020gmf}. The phase shift $\delta_1^1(s)$ is expressed in terms of the pion decay constant $F$ in the chiral limit, the pion mass $\mpi$ (including quark-mass renormalization), and a set of low-energy constants (LECs): at next-to-leading order (NLO) $l_2^r - 2l_1^r$, at next-to-next-to-leading order (NNLO) $l_{1,2,3}^r$, $r_{a,b,c}$, and, in both cases, potentially $l_4^r$ (plus $r_F^r$ at NNLO) to convert $F$ to the physical-point $F_\pi$. Here, we will illustrate the resulting quark-mass dependence using the lattice results for $\pi\pi$ scattering from Ref.~\cite{Andersen:2018mau}, but these LECs could also become free parameters of the lattice analysis. As final ingredient we need the quark-mass dependence of $\langle r_\pi^2\rangle$, which is also known at two-loop order~\cite{Bijnens:1998fm}
\begin{align}
\label{rpi2}
\langle r_\pi^2\rangle&=\frac{1}{16\pi^2 F^2}\bigg[R_4+\frac{\mpi^2}{16\pi^2 F^2}R_6\bigg],\\
R_4&=-96\pi^2l_6^r-1-L,\qquad L=\log\frac{\mpi^2}{\mu^2},\notag\\
R_6&=6(16\pi^2)^2r_{V1}^r+\frac{52\pi^2-181}{48}+\bigg[\frac{19}{6}-96\pi^2\big(2l_1^r-l_2^r\big)\bigg]L.\notag
\end{align}
At NLO the only new LEC, $l_6^r$, is determined from the physical-point $\langle r_\pi^2\rangle=0.429(4)\fm^2$~\cite{Colangelo:2018mtw},\footnote{This value is in agreement with Refs.~\cite{Amendolia:1986wj,Hanhart:2016pcd,Ananthanarayan:2017efc,Zyla:2020zbs}, and, at the quoted precision, is insensitive to $\rho$--$\omega$ mixing, whose relative effect is suppressed by $\epsilon_\omega\sim 2\times 10^{-3}$.} while the quark-mass dependence of the Omn\`es function and its derivative is given by the IAM. At NNLO a new LEC, $r_{V1}^r$, enters, as discussed in more detail below.

\section{Chiral extrapolation of $I=1$ contribution}
\label{sec:extrapolation}

As phenomenological reference point we start from~\cite{Colangelo:2018mtw}
\beq
a_\mu^\text{HVP}[\pi\pi, \leq1\GeV]=494.8(1.4)(2.1)\times 10^{-10},
\eeq
which gives the two-pion contribution to Eq.~\eqref{amu_HVP} up to a cutoff $\Lambda=1\GeV$ and includes final-state radiation in the point-like approximation. Numerically, this dominant, infrared-enhanced contribution increases the HVP integral by $4.2\times 10^{-10}$~\cite{Moussallam:2013una}. In addition, we need to remove the impact of $\rho$--$\omega$ mixing as the second important isospin-breaking effect, which can be done by setting the corresponding mixing parameter $\epsilon_\omega$ in $G_\omega(s)$ to zero, amounting to a shift of $4.3\times 10^{-10}$. In total, we then arrive at 
\begin{align}
\label{HVP_reference}
\bar a_\mu^\text{HVP}[\pi\pi, \leq1\GeV]&\equiv a_\mu^\text{HVP}[\pi\pi, \leq1\GeV]\big|_\text{no FSR, $\epsilon_\omega=0$}\notag\\
&=486.3(1.4)(2.1)\times 10^{-10}
\end{align}
for the two-pion contribution to $a_\mu^\text{HVP}[ud, I=1]$. As a first step, we may compare to the result if $\delta_1^1$ is solely determined via the IAM fits to the lattice data of Ref.~\cite{Andersen:2018mau} (and the physical pion decay constant $F_\pi=92.28(10)\MeV$~\cite{Zyla:2020zbs}), which gives
\begin{align}
\label{HVP_IAM_only}
 \bar a_\mu^\text{HVP}[\pi\pi, \leq1\GeV]\big|^\text{NLO}_{\scalebox{0.7}{\cite{Andersen:2018mau}}}&=458.6(1.9)(14.9)(7.2)\times 10^{-10},\notag\\
 \bar a_\mu^\text{HVP}[\pi\pi, \leq1\GeV]\big|^\text{NNLO}_{\scalebox{0.7}{\cite{Andersen:2018mau}}}&=508.0(28.8)(1.6)(8.5)\times 10^{-10},
\end{align}
where the first error derives from the fit parameters, the second one gives the truncation error in the chiral expansion, estimated for an observable $X$ as~\cite{Niehus:2020gmf,Epelbaum:2014efa}
\begin{align}
    \Delta X_\text{NLO} &= \alpha X_\text{NLO},\qquad \alpha=\frac{\mpi^2}{M_\rho^2},\notag\\
    \Delta X_\text{NNLO} &= \max \left\{\alpha^2X_\text{NLO}, \alpha\left\vert X_\text{NLO} - X_\text{NNLO}\right\vert\right\},\label{eq:truncation}
\end{align}
and the third one propagates the uncertainty in $\langle r_\pi^2\rangle$. The level of agreement between Eqs.~\eqref{HVP_reference} and~\eqref{HVP_IAM_only} reflects the extent to which the extrapolations of the lattice fits to the physical point via the one- and two-loop IAM reproduce the physical phase shift, see Fig.~1 in Ref.~\cite{Niehus:2020gmf}. 

Next, we consider a variant of the IAM fits that includes the physical $\delta_1^1$ from Ref.~\cite{Colangelo:2018mtw} (using $20$ equidistant data points between $0.35\GeV$ and $1.15\GeV$~\cite{SM}). As expected, this reduces the uncertainties substantially 
\begin{align}
\label{HVP1GeV}
 \bar a_\mu^\text{HVP}[\pi\pi, \leq1\GeV]\big|^\text{NLO}_{\scalebox{0.7}{\cite{Andersen:2018mau,Colangelo:2018mtw}}}&=460.4(0.3)(14.9)(7.2)\times 10^{-10},\notag\\
 \bar a_\mu^\text{HVP}[\pi\pi, \leq1\GeV]\big|^\text{NNLO}_{\scalebox{0.7}{\cite{Andersen:2018mau,Colangelo:2018mtw}}}&=482.4(0.1)(0.7)(8.0)\times 10^{-10},
\end{align}
especially in the two-loop fit. In this case, the central value moves close to Eq.~\eqref{HVP_reference}, mainly, because 
the functional form has the necessary freedom to reconcile the expected asymptotic behavior of the phase shift $\delta_1^1\overset{s\to\infty}{\to} \pi$ with the resonant line shape of the $\rho(770)$.  The remaining uncertainty originates from the input for the pion
charge radius, which in turn is dominated by inelastic
effects.\footnote{The value $\langle r_\pi^2\rangle=0.429(4)\fm^2$~\cite{Colangelo:2018mtw} is derived from fits to $e^+e^-\to 2\pi$ data via a sum rule, which implies a sensitivity to $\Im F_\pi^V(s)$ beyond $1\GeV$, where inelastic effects become important.}   Accordingly, the IAM representation reproduces the full result up
to the level at which uncertainties from inelastic channels begin to
matter, but provides a reliable implementation of the elastic $\pi\pi$
effects. This points the way towards the application in the chiral
extrapolation of lattice HVP results: the pion-mass dependence of the
$\pi\pi$ physics can be controlled with the IAM, and only the estimate of
the pion-mass dependence of inelastic effects needs to rely on a
parameterization that is not controlled by effective field theory.  

For larger-than-physical pion masses the $\rho(770)$ resonance also moves
higher in energy, so that a fixed cutoff at $\Lambda=1\GeV$ may not be equally well motivated for all pion masses. With the IAM representation successfully benchmarked against phenomenology, we will thus take $\Lambda\to\infty$ in the following, which changes Eq.~\eqref{HVP1GeV} to
\begin{align}
\label{HVPinfty}
 \bar a_\mu^\text{HVP}[\pi\pi]\big|^\text{NLO}_{\scalebox{0.7}{\cite{Andersen:2018mau,Colangelo:2018mtw}}}&=468.8(0.3)(15.2)(7.6)\times 10^{-10},\notag\\
 \bar a_\mu^\text{HVP}[\pi\pi]\big|^\text{NNLO}_{\scalebox{0.7}{\cite{Andersen:2018mau,Colangelo:2018mtw}}}&=490.8(0.1)(0.7)(8.4)\times 10^{-10},
\end{align}
as starting point for a study of the pion-mass dependence. 
The increase of $8.4\times 10^{-10}$ from Eq.~\eqref{HVP1GeV} to Eq.~\eqref{HVPinfty} is slightly smaller than the $11.8\times 10^{-10}$ obtained when extrapolating the central fit from Eq.~\eqref{HVP_reference} to $\Lambda=\infty$. In both cases, the effect is close to $a_\mu^\text{HVP}[\pi\pi,[1,1.8]\GeV]=10.4\times 10^{-11}$~\cite{Davier:2019can,Keshavarzi:2019abf,Aoyama:2020ynm}, but we stress that this extrapolation is not constrained by $e^+e^-\to 2\pi$ data, and, in the case of the IAM representation, simply serves as a convenient reference point compared to which we will study the quark-mass dependence. In practice, this subsumes some inelastic effects, but their contribution will need to be included as an additional term in any case, see Sec.~\ref{sec:strategies}.

  \begin{figure}[t]
	\includegraphics[width=\linewidth]{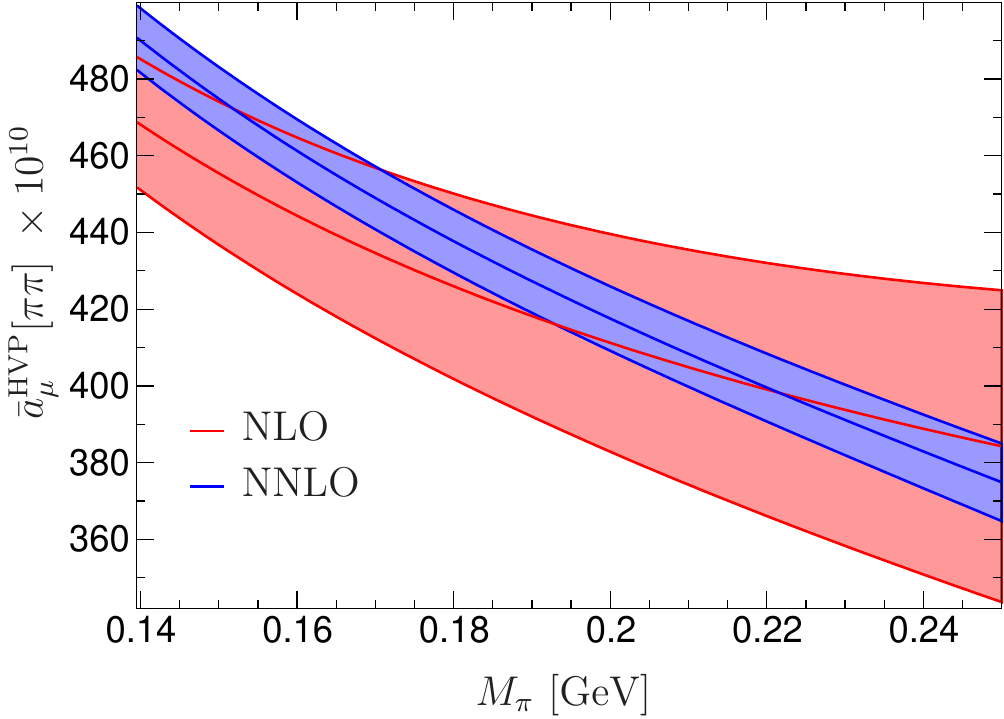}
	\caption{Pion-mass dependence of $\bar a_\mu^\text{HVP}[\pi\pi]$ from the NLO (red) and NNLO (blue) IAM, with parameters determined as described in the main text. The NNLO band is dominated by the uncertainty in the input for $\langle r_\pi^2\rangle$, the NLO band by the truncation of the chiral expansion.} 
	\label{fig:HVP_pion_mass}
\end{figure}

To obtain the pion-mass dependence of $\bar a_\mu^\text{HVP}[\pi\pi]$, we
need to determine the free parameters in Eq.~\eqref{rpi2}. At NLO, only
$l_6^r$ enters, which is then eliminated by imposing $\langle
r_\pi^2\rangle$ at the physical point. At NNLO, however, a new LEC
($r_{V1}^r$) arises, which describes the quark-mass dependence of $\langle
r_\pi^2\rangle$ and is therefore essentially inaccessible to phenomenology.\footnote{Separating $r_{V1}^r$ and $l_6^r$ from the momentum dependence of $F_\pi^V(s)$ is possible, in principle, indirectly via loop effects, but comes at the cost of introducing difficult-to-control systematic errors especially in the determination of $l_6^r$.}  
Instead, we turn to the estimate of $r_{V1}^r$ via resonance saturation~\cite{Ecker:1989yg,Ecker:1988te,Bijnens:1998fm}
\beq
r_{V1}^r=\frac{2\sqrt{2} f_\chi f_V F_\pi^2}{M_V^2},
\eeq
with parameters that can be determined from $\rho\to e^+e^-$, $\rho\to \pi\pi$, and $K^*\to K\pi$
\begin{align}
 \Gamma[\rho\to e^+e^-]&=\frac{(4\pi\alpha)^2M_\rho f_V^2}{12\pi},\\ 
 \Gamma[\rho\to \pi\pi]&=\frac{M_\rho^2\big(M_\rho^2-4\mpi^2\big)^{3/2}}{48\pi F_\pi^4}\bigg(g_V+2\sqrt{2} f_\chi \frac{2\mpi^2}{M_\rho^2}\bigg)^2,\notag\\
 \Gamma[K^*\to K\pi]&=\frac{\lambda^{3/2}\big(M_{K^*}^2,M_K^2,\mpi^2\big)}{48\pi F_K^4 M_{K^*}}\bigg(g_V+2\sqrt{2} f_\chi \frac{\mpi^2+M_K^2}{M_{K^*}^2}\bigg)^2,\notag
\end{align}
where $\lambda(a,b,c)=a^2+b^2+c^2-2(ab+ac+bc)$. Numerically, we find 
\beq
f_V=0.20,\qquad g_V=0.084, \qquad f_\chi=2.5\times 10^{-3},
\eeq
and thus 
\beq
r_{V1}^r=2.0\times 10^{-5},
\label{rV1}
\eeq
where the difference to Ref.~\cite{Bijnens:1998fm} originates from using the kaon decay constant $F_K$ instead of $F_\pi$ in $\Gamma[K^*\to K\pi]$ to minimize SU(3) breaking effects. 
Alternatively, and in the future hopefully more reliably, one could constrain the pion-mass dependence of $\langle r_\pi^2\rangle$ directly from lattice QCD~\cite{Feng:2019geu,Wang:2020nbf}, e.g., at $\mpi=340.9\MeV$ ChPT predicts
\begin{align}
\langle r_\pi^2\rangle\big|_{\mpi=340.9\MeV}^\text{NLO}&=0.373(4)(72)\fm^2,\notag\\
\langle r_\pi^2\rangle\big|_{\mpi=340.9\MeV}^\text{NNLO}&=0.350(4)(13)(7)\fm^2,
\label{rpiChPT}
\end{align}
where the first error is propagated from the physical-point $\langle r_\pi^2\rangle$, the second one estimates the truncation error by the prescription in Eq.~\eqref{eq:truncation}, and the third one arises when assigning a $100\%$ uncertainty to Eq.~\eqref{rV1}. In comparison, Ref.~\cite{Feng:2019geu} quotes 
\beq
\langle r_\pi^2\rangle\big|_{\mpi=340.9\MeV}^\text{\cite{Feng:2019geu}}=0.3485(27)\fm^2,
\eeq
in good agreement with Eq.~\eqref{rpiChPT}. We will thus continue to use Eq.~\eqref{rV1} in the following (including a $100\%$ uncertainty). 

\begin{figure}[t]
	\includegraphics[width=\linewidth]{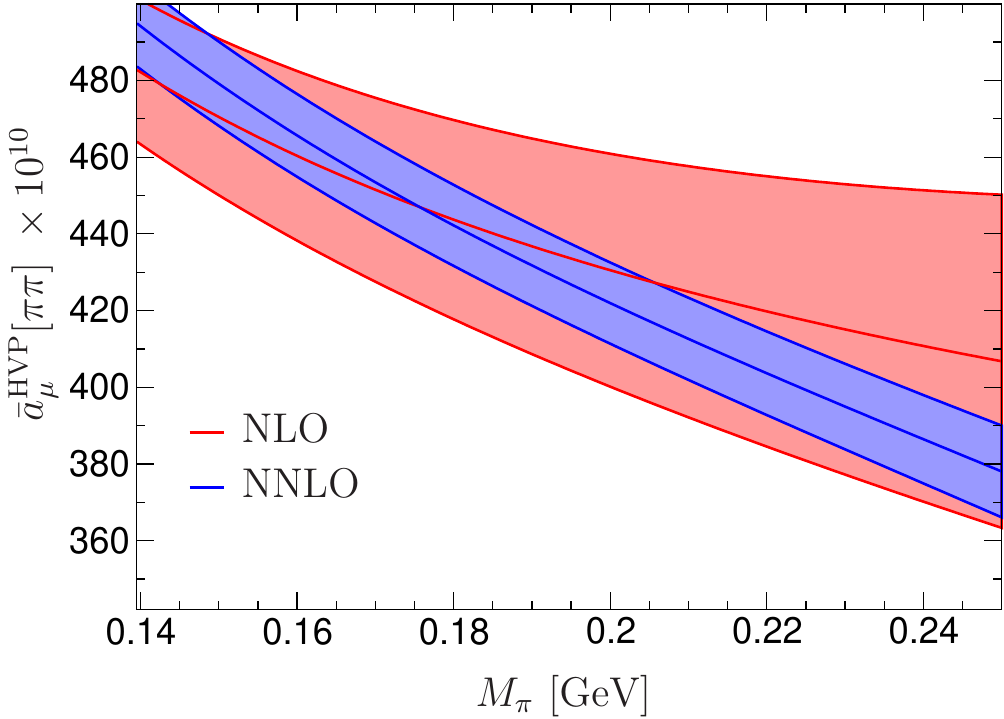}
	\caption{Same as Fig.~\ref{fig:HVP_pion_mass}, but using a conformal polynomial in Eq.~\eqref{VFF_isospin_limit}.} 
	\label{fig:HVP_pion_mass_conf}
\end{figure}

The resulting pion-mass dependence of $\bar a_\mu^\text{HVP}[\pi\pi]$ is shown in Fig.~\ref{fig:HVP_pion_mass}. The NLO band is dominated by the truncation of the chiral expansion, while at NNLO this error stays below $5\times 10^{-10}$ up to $\mpi=0.25\GeV$, with the main effect thus from the uncertainty in the physical $\langle r_\pi^2\rangle$ (and, for higher pion masses, increasingly $r_{V1}^r$). In particular, we observe good consistency between the NLO and NNLO trajectories.

We also considered a variant in which the polynomial in Eq.~\eqref{VFF_isospin_limit} is replaced in favor of a conformal polynomial,
\beq
G_\text{in}(s)=1+\sum_{i=1}^2 c_i\big([z(s)]^i-[z(0)]^i\big),
\eeq
with conformal variable
\beq
z(s)=\frac{\sqrt{s_\text{in}-s_c}-\sqrt{s_\text{in}-s}}{\sqrt{s_\text{in}-s_c}+\sqrt{s_\text{in}-s}},\qquad s_\text{in}=(\mpi+M_\omega)^2,
\eeq
furthermore $s_c=-1\GeV^2$, $c_1=-2c_2$ to remove the $S$-wave singularities, and the remaining parameter again determined via $\langle r_\pi^2\rangle$. The pion-mass dependence of $M_\omega$ is taken from Ref.~\cite{Dax:2018rvs}. The resulting bands, shown in Fig.~\ref{fig:HVP_pion_mass_conf} on the same scale as in Fig.~\ref{fig:HVP_pion_mass}, are well consistent with the parameterization in terms of a linear polynomial, especially the NNLO result is very stable under the change of parameterization. In both cases, one could also include a second free parameter in the (conformal) polynomial, to be identified with the curvature $c_\pi=\frac{1}{2}\frac{d^2F_\pi^V(s)}{ds^2}\big|_{s=0}$, as the only new LEC, $r_{V2}$, would be determined by $c_\pi$ at the physical point. 

\section{Space-like formulation}
\label{sec:spacelike}

\begin{figure}[t]
	\includegraphics[width=\linewidth]{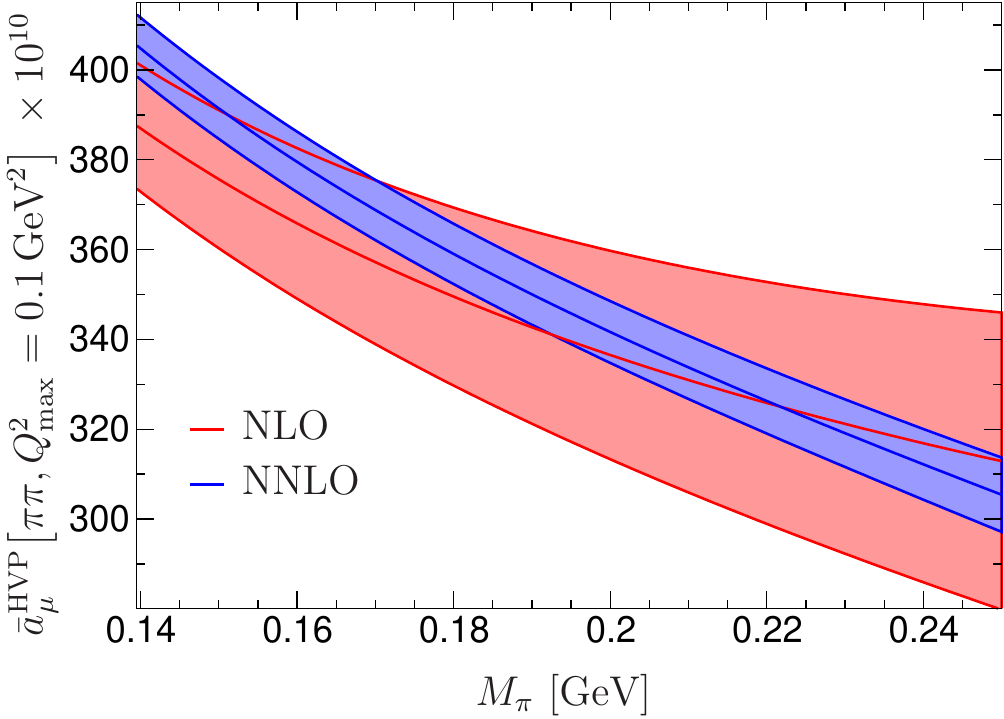}
	\caption{Same as Fig.~\ref{fig:HVP_pion_mass}, but including a space-like cutoff at $Q_\text{max}^2=0.1\GeV^2$.} 
	\label{fig:HVP_pion_mass_Qmax}
\end{figure}

The results shown so far do not indicate any conceptual issues with the pion-mass dependence at least of the two-pion contribution, in contrast to the 
negative conclusions
reached in Ref.~\cite{Golterman:2017njs}. In order to better understand the relation between the two approaches, we will now consider the space-like HVP master
formula, as it was used in Ref.~\cite{Golterman:2017njs} (after
applying a cutoff $Q_\text{max}$). Starting
from~\cite{Lautrup:1971yp,Blum:2002ii} 
\beq
a_\mu^\text{HVP}=-4\alpha^2\int_0^{Q_\text{max}^2}\frac{dQ^2}{Q^2}w\big(Q^2\big)
\bar\Pi\big(-Q^2\big), 
\eeq 
with the subtracted vacuum-polarization function
$\bar \Pi(s)=\Pi(s)-\Pi(0)$ and weight function $w(Q^2)$, the $\pi\pi$
contribution can be evaluated by inserting a dispersion relation for $\bar
\Pi(s)$ and retaining the imaginary part produced by $\pi\pi$ intermediate
states. In the end, this leads to a modification of the time-like master
formula~\eqref{amu_HVP} by a weight function \beq
\theta\big(s,Q_\text{max}^2\big)=\frac{\int_0^{Q_\text{max}^2}dQ^2\frac{w(Q^2)}{s+Q^2}}{\int_0^{\infty}dQ^2\frac{w(Q^2)}{s+Q^2}},
\eeq reminiscent of the definition of Euclidean time
windows~\cite{Blum:2018mom}. The result for $Q_\text{max}^2=0.1\GeV^2$
shown in Fig.~\ref{fig:HVP_pion_mass_Qmax} looks very similar to
Fig.~\ref{fig:HVP_pion_mass}, only with a lower overall scale given that
part of the integral has been removed by the space-like cutoff. In
Ref.~\cite{Golterman:2017njs} this cutoff is required because ChPT is used
directly for the vector correlator~\cite{Golowich:1995kd,Amoros:1999dp},
which does include effects beyond the $\pi\pi$ channel, but restricts the
domain of validity in the momentum integral. The approach we are suggesting
here does not rely on any cutoffs and amounts to instead concentrating on
the $\pi\pi$ channel (including some inelastic effects via the pion charge
radius), since the pion-mass dependence can then be controlled via the IAM,
with only the remainder to be described by simpler parameterizations.

\begin{figure}[t]
	\includegraphics[width=0.495\linewidth]{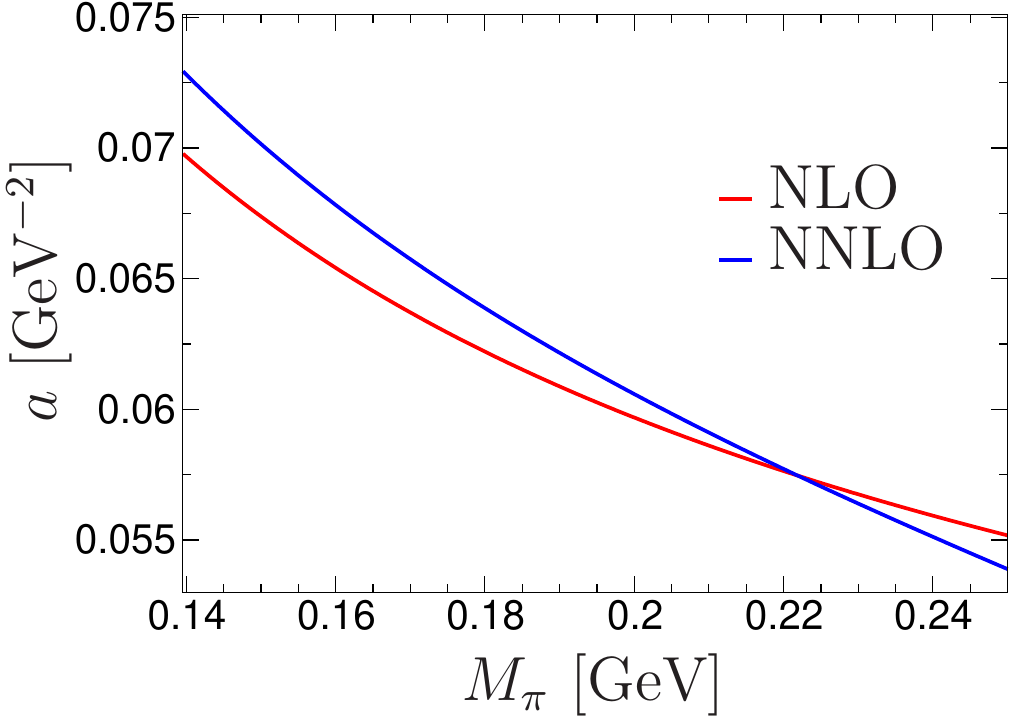}
	\includegraphics[width=0.495\linewidth]{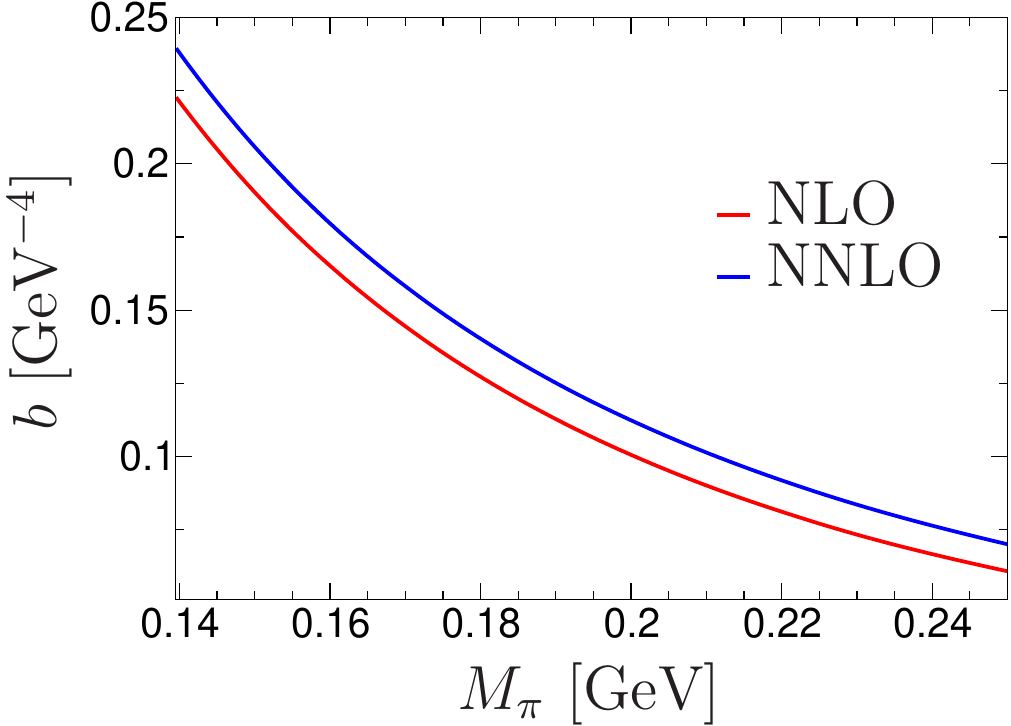}\\
	\includegraphics[width=0.495\linewidth]{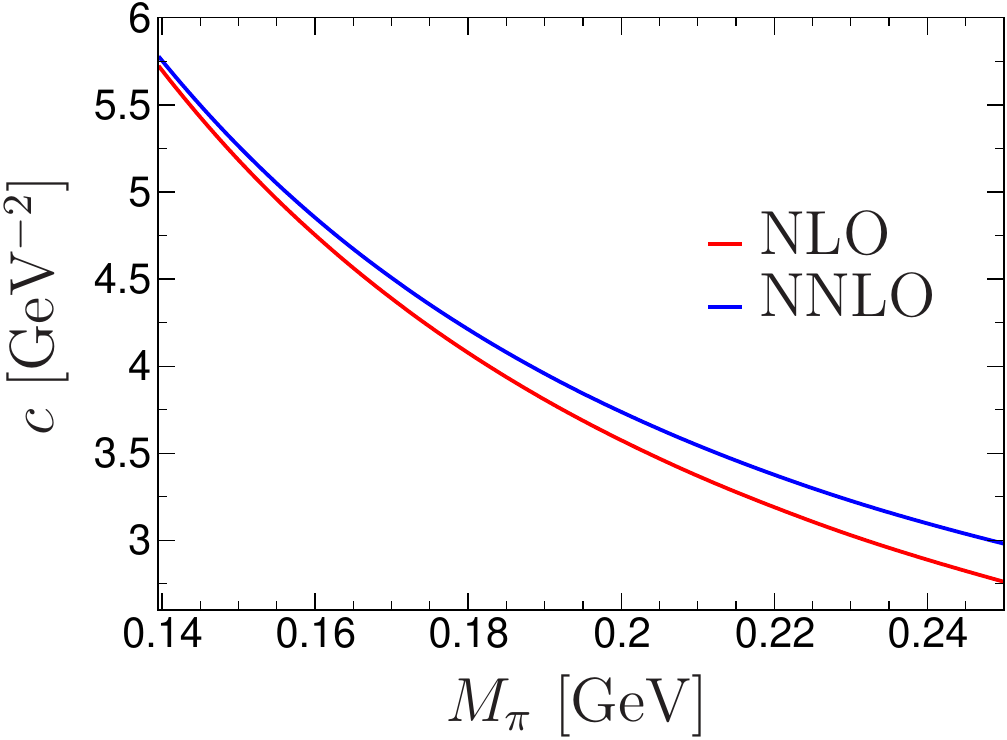}
	\includegraphics[width=0.495\linewidth]{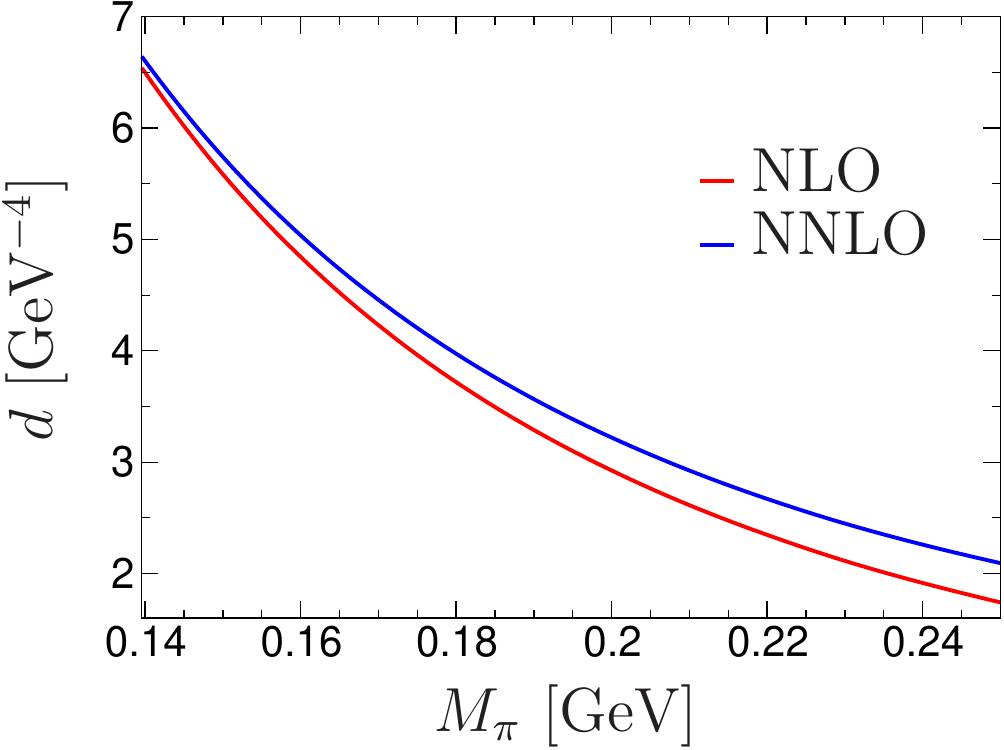}
	\caption{Pion-mass dependence of the fit coefficients $\{a,b,c,d\}$ in Eq.~\eqref{PiQ2} as predicted by the NLO (red) and NNLO (blue) IAM, with the same input for the VFF as in Fig.~\ref{fig:HVP_pion_mass}.} 
	\label{fig:abcd}
\end{figure}

In the space-like region, the integrand $\bar \Pi(-Q^2)$ becomes
sufficiently smooth that simple descriptions in terms of a few parameters
become possible, e.g., we have checked that the $\pi\pi$ contribution to
$\bar \Pi(-Q^2)$ can be represented by the ansatz 
\beq
\label{PiQ2}
\frac{\bar \Pi(-Q^2)}{Q^2}=\frac{a+b Q^2}{1+c Q^2+d Q^4},
\eeq
with an accuracy below $10^{-3}$ for $\mpi\in[0.14,0.25]\GeV$ and $Q^2\in[0,10]\GeV^2$.  We can thus express the IAM prediction for the pion-mass dependence of $\bar \Pi(-Q^2)$ in terms of the fit coefficients in Eq.~\eqref{PiQ2}, see Fig.~\ref{fig:abcd}. We tried several ans\"atze
\begin{align}
 f_1(\mpi^2)&=x + y \mpi^2 + z \mpi^4,\notag\\
 f_2(\mpi^2)&=x \log \mpi^2 + y + z \mpi^2,\notag\\
 f_3(\mpi^2)&=\frac{x}{\mpi^2}+y + z \mpi^2,\notag\\
 f_4(\mpi^2)&=\frac{x}{\mpi^2}+y \log \mpi^2 + z,
\end{align}
cf.\ also Ref.~\cite{Golterman:2017njs},
for the pion-mass dependence of $\{a,b,c,d\}$ at NLO and NNLO, with the result that $f_1$ and in most cases $f_2$ are (strongly) disfavored, indicating that an $\mpi^{-2}$ term is necessary. The NLO result for $a$ displays some preference for an additional $\log \mpi^2$ term, but for all other coefficients, and also for $a$ at NNLO, $f_3$ and $f_4$ describe the pion-mass dependence from the IAM equally well.  Extending these empirical fits to include pion masses below the physical point increases the sensitivity to the infrared singularities, with results that suggest the presence of a $\log \mpi^2$ in addition to the $\mpi^{-2}$ term.  

In principle, a similar strategy could also be pursued for the integrand $G(t)$ in the time-momentum representation~\eqref{timemomentum}, but due to its more complicated behavior an accurate description in terms of few parameters in analogy to Eq.~\eqref{PiQ2} is difficult to find. In contrast, the pion-mass dependence of $\bar a_\mu^\text{HVP}[\pi\pi]$ is sufficiently smooth to be described by $f_{2\text{--}4}$ with an error below $1\times 10^{-10}$ for $\mpi\in[0.14,0.25]\GeV$ when fitting the central curves. Again, these empirical fits favor the presence of an $\mpi^{-2}$ term in the extrapolation, and when including pion masses below the physical point we also see indications for the presence of an additional $\log \mpi^2$ singularity.  
We emphasize that these findings are purely empirical, to describe the IAM results in a finite range of $\mpi$, and we do not claim that either fit function represent an analytic approximation to the full IAM.

\section{Possible implementation strategies}
\label{sec:strategies}

The above results suggest two main strategic approaches to the chiral
extrapolation of lattice results for HVP, each of which could then be
implemented following different variants. The first would explicitly rely
on the description of the dominant $\pi\pi$ contribution in terms of the
$P$-wave phase shift, whose quark-mass dependence has been shown to be well
described by the IAM~\cite{Niehus:2020gmf}, and use the LECs appearing in that
description directly as fit parameters. Possible variants for the implementation of this strategy include:
\begin{enumerate}
 \item In the most constrained scenario, the free parameters of the IAM could be taken from an independent lattice calculation of the $\pi\pi$ $P$-wave and the pion decay constant, leaving only the parameter $\beta$ as a free parameter. In fact, to use the maximum amount of chiral input, the charge radius $\langle r_\pi^2\rangle$ at the physical point could be identified as fit parameter (instead of $\beta$), given that this input yields the dominant uncertainty in predicting the pion-mass dependence of $\bar a_\mu^\text{HVP}[\pi\pi]$, and further lattice input could constrain $r_{V1}^r$. 
 \item Relaxing chiral constraints, (some of) the LECs could be left free as additional fit parameters. 
 \item Since $\bar a_\mu^\text{HVP}[\pi\pi]$ does not saturate
   $a_\mu^\text{HVP}[ud,I=1]$, it needs to be supplemented by an additional
   term, in the simplest case
   $a_\mu^\text{HVP}[ud,I=1,\text{non-}\pi\pi]=\zeta+\mpi^2\xi$, without
   further chiral constraints on the parameters. It is clear that higher
   intermediate states can only affect the singularity structure of
   $a_\mu^\text{HVP}$ towards the chiral limit at higher chiral orders. 
   We thus consider it highly unlikely that these higher intermediate states would affect the parameterization we adopted to describe the chiral extrapolation in any perceivable way, so that a simple phenomenological description should be sufficient for any practical purposes.
\end{enumerate}
We stress that the bands in Fig.~\ref{fig:HVP_pion_mass} appear broad
compared to the sub-percent target precision, but also that, crucially, at
NNLO the uncertainty is dominated by external input quantities (most
notably $\langle r^2_\pi\rangle$), with the chiral convergence well under
control. This implies that in the opposite direction, extrapolating HVP
values at larger-than-physical pion masses towards the physical point, the
intrinsic uncertainty to be assigned to the $\pi\pi$ component should be
small, with the dominant sources of uncertainty likely from the required
LECs and the chiral extrapolation of the non-$\pi\pi$ contribution. In
particular, the LECs that enter the extrapolation can be constrained from
independent lattice calculations for $\delta_1^1$, $F_\pi$, and $\langle
r_\pi^2\rangle$.

The second strategic approach to the chiral extrapolation of the HVP
contribution only indirectly relies on the description of the two-pion
contribution in terms of the $P$-wave phase shift. The latter is used only
to show that in the space-like region the polarization function due to the
$\pi\pi$ contribution is sufficiently smooth to allow for an accurate
representation in terms of just a few fit parameters, whose quark-mass
dependence is again well described by simple parameterizations. One would
then adopt one (or a few) of the parameterizations proposed and tested here
and fit its parameters to the lattice data to perform the chiral
extrapolation. Here the possible variants would depend on the amount and
precision of lattice data and would boil down to choosing among the
different possible parameterizations discussed here or on further
refinements thereof.

\section{Conclusions}
\label{sec:conclusions}

In this Letter we studied potential strategies to control the quark-mass dependence of the HVP contribution to the anomalous magnetic moment of the muon based on effective field theory. In particular, we focused on the $I=1$ component of the isospin-symmetric $ud$ correlator, which receives its by far dominant contribution from $\pi\pi$ intermediate states, with $4\pi$ and other non-$\pi\pi$ contributions appreciably suppressed. A direct application of ChPT is not possible due to the limited range of convergence~\cite{Golterman:2017njs}, ultimately, because information on the $\rho$ meson needs to be provided. Here, we argued that this can be achieved based on the IAM, with one- and two-loop implementations allowing one to verify the chiral convergence and thus assess the corresponding systematic uncertainties. 

To illustrate the formalism, we addressed the opposite problem, to predict the quark-mass dependence starting from the physical point, with the main result shown in Fig.~\ref{fig:HVP_pion_mass}, based on input from the $\pi\pi$ $P$-wave phase shift $\delta_1^1$ and the pion charge radius $\langle r_\pi^2\rangle$ at the physical point~\cite{Colangelo:2018mtw}, combined with a lattice calculation of $\delta_1^1$ and the pion decay constant $F_\pi$~\cite{Andersen:2018mau} to determine the ChPT parameters. We found that the IAM representation indeed allows one to reproduce the expected HVP value, and suggested strategies how it could be used to constrain the chiral extrapolation of HVP calculations in lattice QCD performed at larger-than-physical quark masses. 

Moreover, we found that in the space-like region the IAM results are
sufficiently smooth that the correlator can be described by simple
parameterizations. We could successfully reproduce their quark-mass dependence by
simple fit functions, with the result that the presence of an infrared
singularity as strong as $\mpi^{-2}$ seems to be preferred empirically.
Altogether, we see no conceptual issues in controlling the quark-mass
dependence of HVP in lattice calculations along these lines using effective
field theory, with several opportunities to constrain the extrapolation
with independent lattice input. Of course, the precision that can be
reached in extrapolations to the physical point, or even interpolations
around it, will strongly depend on the details of each lattice calculation,
but the possibility to achieve high precision even if working away from the
physical point and reaching it by an extrapolation stays open.

\section*{Acknowledgments}

We acknowledge valuable discussions at~\cite{KEK} that triggered the present write-up, and thank Marco C\`e, Aida El-Khadra, and Maarten Golterman for subsequent discussions. 
Financial support by the Bonn--Cologne Graduate School of Physics and Astronomy (BCGS), the DFG (CRC 110, ``Symmetries and the Emergence of Structure in QCD''), and the Swiss National Science Foundation, under Project Nos.\ PCEFP2\_181117 (M.H.) and PZ00P2\_174228 (J.R.E.) and Grant No.\
200020\_175791 (G.C.), is gratefully acknowledged. 
  
\bibliographystyle{apsrev4-1_mod}
\balance
\biboptions{sort&compress}
\bibliography{AMM}

\end{document}